\documentclass[referee]{raa}
\usepackage{graphicx,times}
\usepackage{natbib}
\usepackage{amssymb,amsmath}
\bibpunct{(}{)}{;}{a}{}{,}

\usepackage[a4paper=true,dvipdfm=true,pagebackref=true]{hyperref}
\hypersetup{pdftitle = The title of my PDF, pdfauthor = My name, pdfsubject= The subject, pdfkeywords = keyword1 keyword2 keyword3}
\hypersetup{colorlinks = true, linkcolor = green, anchorcolor = red, citecolor = blue, filecolor = red, pagecolor = red, urlcolor = red}

\usepackage{color}

\begin{document}

   \title{Morphology and structure of extremely red objects at $z\sim1$ in the CANDELS-COSMOS field
$^*$
\footnotetext{\small $*$ Supported by the National Natural Science Foundation of China.}
}

 \volnopage{ {\bf 2014} Vol.\ {\bf X} No. {\bf XX}, 000--000}
   \setcounter{page}{1}

   \author{Guan-Wen Fang\inst{1,2}, Zhong-Yang Ma\inst{2,3}, Yang Chen\inst{2,3,4}, Xu Kong
      \inst{2,3}   }
   \institute{Institute for Astronomy and History of Science and Technology, Dali University, Dali 671003, China; {\it wen@mail.ustc.edu.cn}\\
	\and
Key Laboratory for Research in Galaxies and Cosmology, The University of Science
and Technology of China, Chinese Academy of Sciences, Hefei, Anhui, 230026, China\\
\and
Center for Astrophysics, University of Science and Technology of China, Hefei 230026, China; {\it xkong@ustc.edu.cn}\\
\and
SISSA, via Bonomea 265, I-34136 Trieste, Italy\\
\vs \no
   {\small Received 2014 ? ?; accepted 2014 ? ?}
}

\abstract{Using high-resolution HST/Wide Field Camera 3 F125W imaging from the CANDELS-COSMOS field, we report the
the structural and morphological properties of Extremely Red Objects (EROs) at $z\sim1$. Based on the UVJ color
criteria, we separate EROs into two types: old passive galaxies (OGs) and dusty star-forming galaxies (DGs).
For a given stellar mass, we find that the mean size of OGs (DGs) is smaller by a factor of $\sim2$ (1.5)
than that of present-day early-type (late-type) galaxies at rest-frame optical wavelength. We derive
the average effective radii of OGs and DGs, corresponding to $2.09\pm1.13$ kpc and $3.27\pm1.14$ kpc, respectively.
Generally, The DGs are heterogeneous, with mixed features including bulges, disks,
and irregular structures, with relatively high $M_{\rm 20}$, small size and low $G$,
while OGs are elliptical-like compact morphologies with lower $M_{\rm 20}$,
larger size and higher $G$, indicating the more concentrated and symmetric spatial
extent of stellar population distribution in OGs than DGs.
The findings imply that OGs and DGs have different evolutionary processes, and the minor
merger scenario is the most likely mechanism for the structural properties of OGs. However,
the size evolution of DGs is possibly due to the secular evolution of galaxies.
}

   \authorrunning{G.-W. Fang et al. }            
   \titlerunning{Morphology and structure of EROs}  
   \maketitle

%
\section{Introduction}
Over the past decade, studies of the $z\sim1-2$ universe have been
revolutionized by the availability of deep
near-infrared (NIR) imaging surveys. One of the primary early results was
the discovery of a population of optically-faint, massive galaxies which
are missed in optical (rest-frame ultraviolet) surveys (Kong et al. 2006).
Extremely Red Objects [EROs with $(I-K)_{\rm Vega}>4$ or $(i-K)_{\rm AB}>2.45$] were first hinted
at by near-infrared selected surveys (Elston et al. 1988, 1989; Hu \& Ridgway 1994; Graham \&
Dey 1996; Dey et al. 1999). They were classified as two different galaxy types using a variety of
observational methods (Wilson et al. 2007; Fang et al. 2009; Kong et al. 2009):
old passive galaxies (OGs) and dusty star-forming galaxies (DGs). As illustrated in the literature
(Stockton et al. 2006; Stern et al. 2006; Conselice et al. 2008; Kong et al. 2009; Kim et al. 2014),
EROs may be the descendants of high-redshift galaxies and progenitors of present-day massive
early-type galaxies. Therefore, EROs are an important population for understanding the formation
and evolution of galaxies.

Utilizing Hubble Space Telescope (HST ACS, WFPC2, and NICMOS) optical and NIR imaging from different surveys, many groups
studied the morphological properties of EROs at $z\sim1$, they found that OGs in the ERO sample have
elliptical-like structures, whereas the morphologies of DGs include disks and irregular systems
(Moriondo et al. 2000; Cimatti et al. 2003; Yan \& Thompson 2003; Giavalisco et al. 2004;
Moustakas et al. 2004; Conselice et al. 2008; Fang et al. 2009; Kong et al. 2009).
Based on a large sample of $\sim5300$ EROs with $K_{\rm AB}\leqslant21.1$ and $(i-K)_{\rm AB}\geqslant2.45$
from the COSMOS field, Kong et al. (2009) found that OGs (48\%) and DGs (52\%) have similar fractions.
Moreover, the brighter ($K$-band magnitude) EROs have stronger clustering amplitude than fainter sources.
The similar findings is also confirmed in Kim et al. (2014).

Half-light radius ($r_{\rm e}$) of galaxies is one of the primary parameters to analyze the
assembly history of galaxy's mass and the galaxy evolutionary paths. For a given stellar mass,
many observations have reported that the size of high-redshift massive galaxies is on average
smaller than that of local counterparts (Daddi et al. 2005; Toft et al. 2005; Trujillo et al. 2006;
Zirm et al. 2007; Cimatti et al. 2008; van Dokkum et al. 2008; Franx et al. 2008; Szomoru et al. 2012;
Gobat et al. 2012; Fan et al. 2013; Patel et al. 2013; Fang et al. 2014; Morishita et al. 2014;
van der Wel et al. 2014). It is difficult to study morphologies of high redshift
galaxies based on their observed optical images, owning to their observed
optical light probes the rest-frame ultraviolet (UV) emission for objects
at $z\geq1$. For instance, pictures of galaxies at $z\geq1.0$ taken with HST WFPC2 and
ACS are all imaged in the rest-frame UV, their apparent morphologies can easily be changed by
patchy dust extinction and star-forming regions. Therefore, it is essential to study
$z\geq1$ galaxies from observed NIR bands, which probe the rest-frame optical morphologies.

In this work, for the first time, we will adopt HST/Wide Field Camera 3
(WFC3) F125W images ($0''.06~{\rm pixel}^{-1}$) to
study the structural and morphological properties of $z\sim1$ EROs at
rest-frame optical wavelength ($\lambda_{\rm rest}\sim6300$~{\AA}). Throughout this paper, we use a
standard cosmology with $\Omega_{\rm M} = 0.3$, $\Omega_\Lambda =0.7$ and a Hubble constant of
$H_{\rm 0}=70$~km s$^{-1}$ Mpc$^{-1}$. At redshift $z\sim1$, $1''.0$ corresponds to 8.0 kpc.
All magnitudes adopt AB system unless otherwise specified.

\section{Observation and data}

The Cosmic Evolution Survey (COSMOS) is a multi-band (from X-ray to radio) survey designed to
probe the formation and evolution of galaxies as a function of redshift and large scale
structure environment, covering an area of $\sim2~{\rm deg^{2}}$ (Scoville et al. 2007).
The more details of multi-band observation and data reduction in the COSMOS field can
be found in McCracken et al. (2012) and Muzzin et al. (2013a). Photometric data
(include redshift $z_{\rm p}$ and stellar mass $M_{\ast}$) we adopt in our study is from
the $K$-selected catalog of the COSMOS/UltraVISTA field provided by Muzzin et al. (2013a).
Meanwhile, we also use HST/WFC3 F125W high-resolution imaging ($0''.06~{\rm pixel}^{-1}$) to
analyze the structural features of EROs in our sample. HST/WFC3 F125W images covers
a total of $\sim$210~\rm arcmin$^2$ in the CANDELS\footnote{Cosmic Assembly Near-IR
Deep Extragalactic Legacy Survey (CANDELS; Grogin et al. 2011 and
Koekemoer et al. 2011)}-COSMOS field, and the $5~\sigma$ point-source detection limit
is 27.0 mag. Further details are in Grogin et al. (2011) for the survey and observational design,
and Koekemoer et al. (2011) for the data products.

\section{Selection and classification of EROs}

Following the photometric technique of EROs we performed in our previous works
(Fang et al. 2009; Kong et al. 2009), we construct a sample of 5241 EROs with
$i-K>2.45$ and $K<21.0$ in the COSMOS field (see Figure 1).
The NIR $K$-band and optical $i$-band data are from VISTA/VIRCAM and
Subaru/SuprimeCam, respectively. In Figure 2, we show redshift ($z_{\rm p}$, panel (a)) and stellar
mass ($M_{\ast}$, panel (b)) distribution of EROs in the COSMOS field. As we saw in Figure 2,
the stellar masses of EROs mainly distribute at $10^{10.3}<M_{\ast}<10^{11.5}$, and
most of them have redshifts at the range $0.8<z_{\rm p}<1.6$. Moreover, EROs represent 87.6\%
of massive galaxies ($M_{\ast}>10^{10.6}{\rm M_{\odot}}$) with $0.9<z_{\rm p}<1.3$ and $K<21.0$,
indicating that the ERO selection criterion is sensitive to select massive galaxies at intermediate redshift.

To investigate the morphological and structural
properties of two primary classes in our EROs sample, separating EROs into
OGs and DGs is necessary. As shown in Figure 3, using the rest-frame UVJ color
criteria defined by Muzzin et al. (2013b), we classify the sample of EROs into
2606 OGs and 2635 DGs, and their fractions correspond to 49.7\% and 50.3\%, respectively.
Our findings are consistent with the results of Mannucci et al. (2002), Giavalisco et al. (2004),
Moustakas et al. (2004), and Kong et al. (2009). Moreover, we also find that
the fraction of DGs in the COSMOS field increases toward fainter
magnitudes: 50.3\% at $K = 21.0$, 52.0\% at $K = 21.5$
, 54.5\% at $K = 22.0$, 56.4\% at $K = 22.5$ and  58.0\% at $K = 23.0$.
The results indicate that DGs become increasingly important at fainter magnitudes.
Figure 4 shows the redshift distribution of OGs (panel (a)) and DGs (panel (b))
in our sample.

\begin{figure*}
\centering
\includegraphics[angle=0,width=0.8\textwidth]{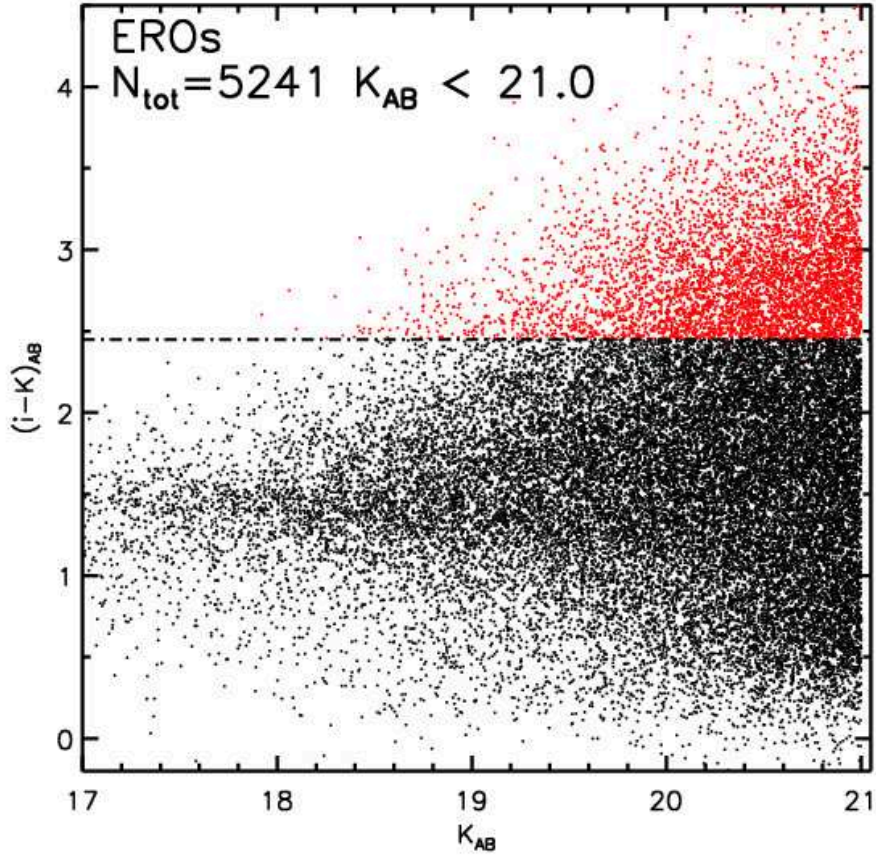}
\caption{Selection of EROs in the COSMOS field. Red dots represent
5241 EROs with $i-K>2.45$ (dot-dashed line) and $K<21.0$.
} \label{fig:eros}
\end{figure*}

\begin{figure*}
\centering
\includegraphics[angle=0,width=0.8\textwidth]{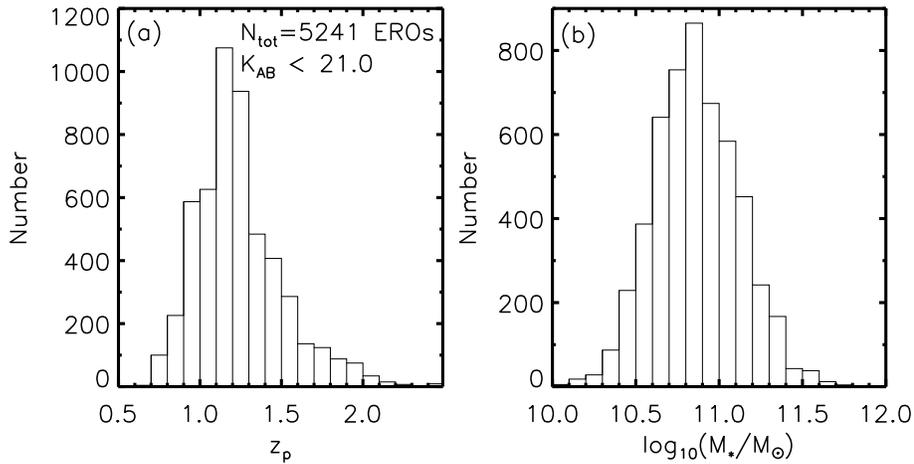}
\caption{Redshift ($z_{\rm p}$) and stellar mass ($M_{\ast}$) histogram of
EROs in the COSMOS field. The left panel (a) is the distribution for $z_{\rm p}$,
and the right panel (b) is the distribution for $M_{\ast}$.
} \label{fig:zmhist}
\end{figure*}

\begin{figure*}
\centering
\includegraphics[angle=0,width=0.8\textwidth]{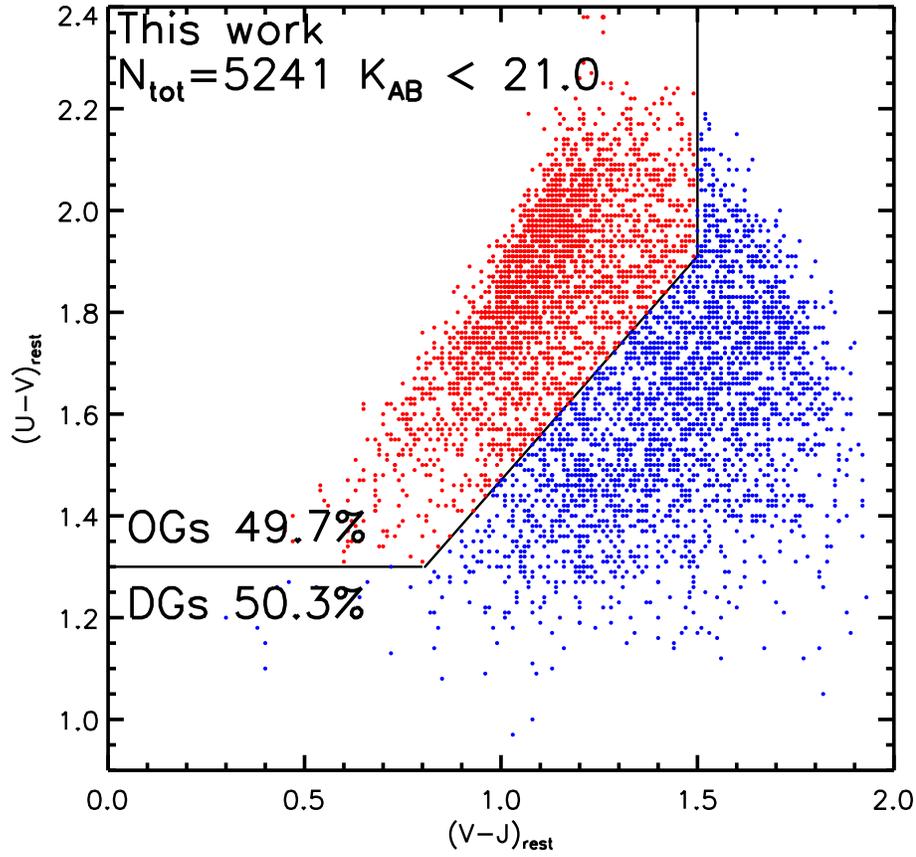}
\caption{Distribution of EROs in the $(U-V)_{\rm rest}$ vs. $(V-J)_{\rm rest}$ diagram.
Solid lines correspond to the color criteria from Muzzin et al. (2013b).
Red and blue dots represent OGs and DGs, respectively.
} \label{fig:class}
\end{figure*}

\begin{figure*}
\centering
\includegraphics[angle=0,width=0.8\textwidth]{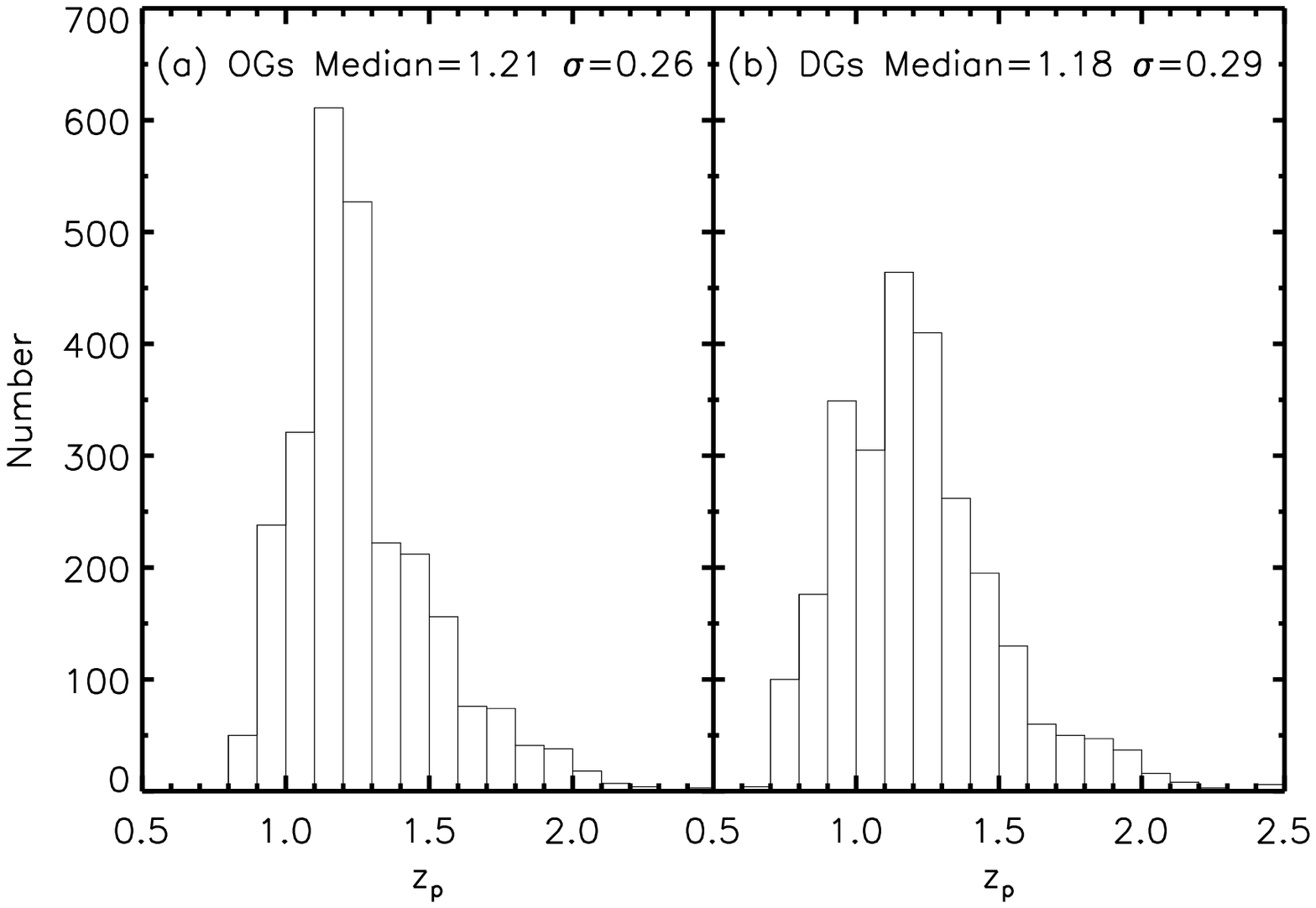}
\caption{Redshift ($z_{\rm p}$) histogram of
EROs in our sample. Panel (a) shows OGs. Panel (b) shows DGs.
} \label{fig:zero}
\end{figure*}

\section{Structures and morphologies of EROs}

The physical sizes of intermediate-redshift massive galaxies can help us to understand the
formation and evolution of local massive galaxies. Within a matched radius of $0''.5$,
we obtain the effective radii of 58 DGs and 55 OGs from the
catalog\footnote{http://www.mpia-hd.mpg.de/homes/vdwel/candels.html} (version 1.0)
provided by van der Wel et al. (2012). The sizes of these EROs are measured using
the HST/WFC3 F125W imaging, and all sources are brighter than
$J({\rm F125W})\sim22.5$. In Figure 5, the size in our EROs sample are
compared to those of $z\sim0.1$ galaxies from Shen et al. (2003). We find that
EROs (include OGs and DGs) at $z\sim1$ follow a clear $r_{\rm e}-M_{\ast}$ relation.
However, most of them have smaller sizes, compared to local counterparts with similar stellar
mass. Moreover, the sizes of DGs are larger than
OGs in general, even in massive systems, but some have very compact structures,
with $r_{\rm e}<1 {\rm kpc}$.

\begin{figure*}
\centering
\includegraphics[angle=0,width=0.85\textwidth]{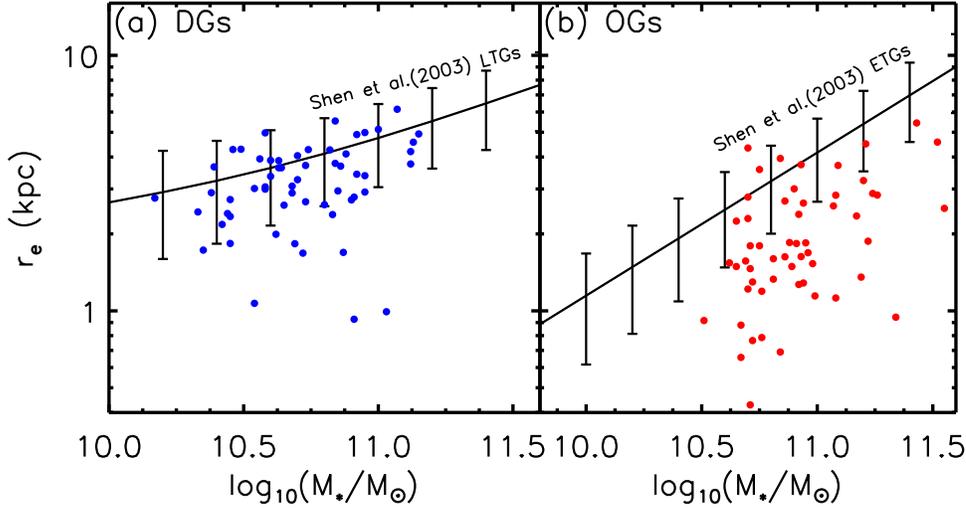}
\caption{Relation of stellar mass ($M_{\ast}$) and size ($r_{\rm e}$) for EROs at $z\sim1.0$.
The solid lines with $1\sigma$ standard error are provided by Shen et al. (2003)
for local late-type and early-type galaxies (LTGs and ETGs).
} \label{fig:rem}
\end{figure*}

To further analyze the evolution of sizes with redshift in our EROs sample, we show
the sizes for OGs ($<r_{\rm e}>=2.09\pm1.13$ kpc and $<z_{\rm p}>=1.31\pm0.23$)
and DGs ($<r_{\rm e}>=3.27\pm1.14$ kpc and $<z_{\rm p}>=1.08\pm0.24$) in Figure 6, respectively.
For comparison with our results, the data of quiescent galaxies (QGs) and star-forming galaxies
(SFGs) from other works are plotted in this figure (Shen et al. 2003; Gobat et al. 2012;
Szomoru et al. 2012; Fan et al. 2013; Patel et al. 2013; Fang et al. 2014; Morishita et al. 2014).
From Figure 6, the mean size of local QGs is two times larger than the mean size of the $z\sim1$
OGs at a fixed stellar mass. For DGs in EROs sample, the effective radius is on average smaller
by a factor of $\sim1.5$ than that of $z\sim0.1$ SFGs with analogous stellar mass.
The smaller sizes and higher masses of EROs in our sample, indicates that EROs have larger
stellar mass surface densities, compared to present-day massive galaxies. By combining with
the results from the literature, we conclude that OGs and DGs have different evolutionary
tracks, the structural properties of OGs are consistent with predictions of the
hierarchical merging models (e.g., dry minor merger), but the size of DGs supports
predictions of the monolithic collapse scenario (e.g., the secular evolution of galaxy).

\begin{figure*}
\centering
\includegraphics[angle=0,width=0.85\textwidth]{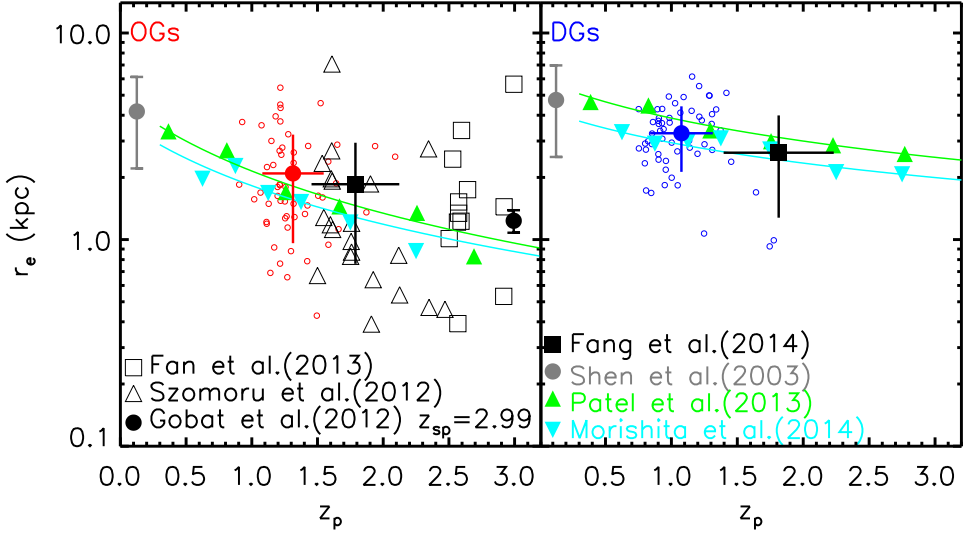}
\caption{Evolution of size with redshift ($z_{\rm p}$) in our EROs sample.
The sizes of QGs and SFGs from the literature are also plotted in this figure.
The mean size of OGs is $2.09\pm1.13$ kpc (red solid circle), while DGs is
$3.27\pm1.14$ kpc (blue solid circle). $Left$: Green and cyan lines
correspond to $r_{\rm e}\propto(1+z)^{-1.16}$ (Patel et al. 2013)
and $r_{\rm e}\propto(1+z)^{-1.06}$ (Morishita et al. 2014), respectively.
$Right$: Green and cyan lines represent $r_{\rm e}\propto(1+z)^{-0.63}$ (Patel et al. 2013)
and $r_{\rm e}\propto(1+z)^{-0.56}$ (Morishita et al. 2014), respectively.
} \label{fig:rez}
\end{figure*}

In order to quantitatively investigate the morphological features of EROs at $z\sim1$,
we measure the morphological parameters of galaxy using HST/WFC3 F125W images.
Such as Gini coefficient ($G$; the relative distribution of the galaxy pixel flux values)
and high moment ($M_{\rm 20}$; the second-order moment of the brightest 20$\%$ of the
galaxy's flux). According to the definition of Abraham et al. (1996)
and Lotz et al. (2004),
\begin{equation}
\label{eq:G} G = \frac{\sum^{N}_{l} (2l-N-1) |F_{l}|}{\overline{F}N(N-1)},
\end{equation}
where $N$ is the total number of pixels in a galaxy, and $\overline{F}$ is the
mean pixel flux of all $F_{l}$ (each pixel flux).
\begin{equation}
\label{eq:M20} M_{\rm 20} = {\rm log}_{\rm 10}(\frac{\sum^{k}_{l=1}M_{l}}{M_{\rm tot}}),
\end{equation}
${\rm where} \sum^{k}_{l=1}F_{l}=0.2F_{\rm tot} {~\rm and~} M_{\rm tot}=\sum^{N}_{l=1}M_{l}$.
Moreover, sort $F_{l}$ by descending order with $|F_{1}|\geqslant|F_{2}|\geqslant\cdot\cdot\cdot\cdot
|F_{k}|\cdot\cdot\cdot\cdot\geqslant|F_{N}|$.
\begin{equation}
\label{eq:G} M_{l} = F_{l}[(x_{l}-x_{o})^2+(y_{l}-y_{o})^2],
\end{equation}
where ($x_{o},~y_{o}$) and ($x_{l},~y_{l}$) represent the galaxy's center and each pixel
position in Cartesian coordinates, respectively.

As shown in Figure 7, OGs in appearance are very similar to
local early-type galaxies, they have higher $G$
and lower $M_{\rm 20}$, compared to DGs. For the morphological properties of
DGs, the majority of them shows diffuse or irregular structures (low $G$ and high $M_{\rm 20}$),
this is similar to late-type galaxies we see today. That indicates less
concentrated and symmetric spatial distribution of the stellar mass of DGs
at $z\sim2$, comparing to OGs. The mean ($M_{\rm 20}$, $G$) values
for DGs are ($-$1.65, 0.58) at $\lambda_{\rm rest}\sim6300$~{\AA}, whereas OGs are
($-$1.81, 0.68). From the morphological analysis of EROs in the CANDELS-COSMOS field, we
conclude that OGs and DGs have different formation modes and the history of mass assembly,
and that the population of Hubble sequence galaxies roughly matches that
of the peculiars sometime between $z = 1-2$.

\begin{figure*}
\centering
\includegraphics[angle=0,width=0.8\textwidth]{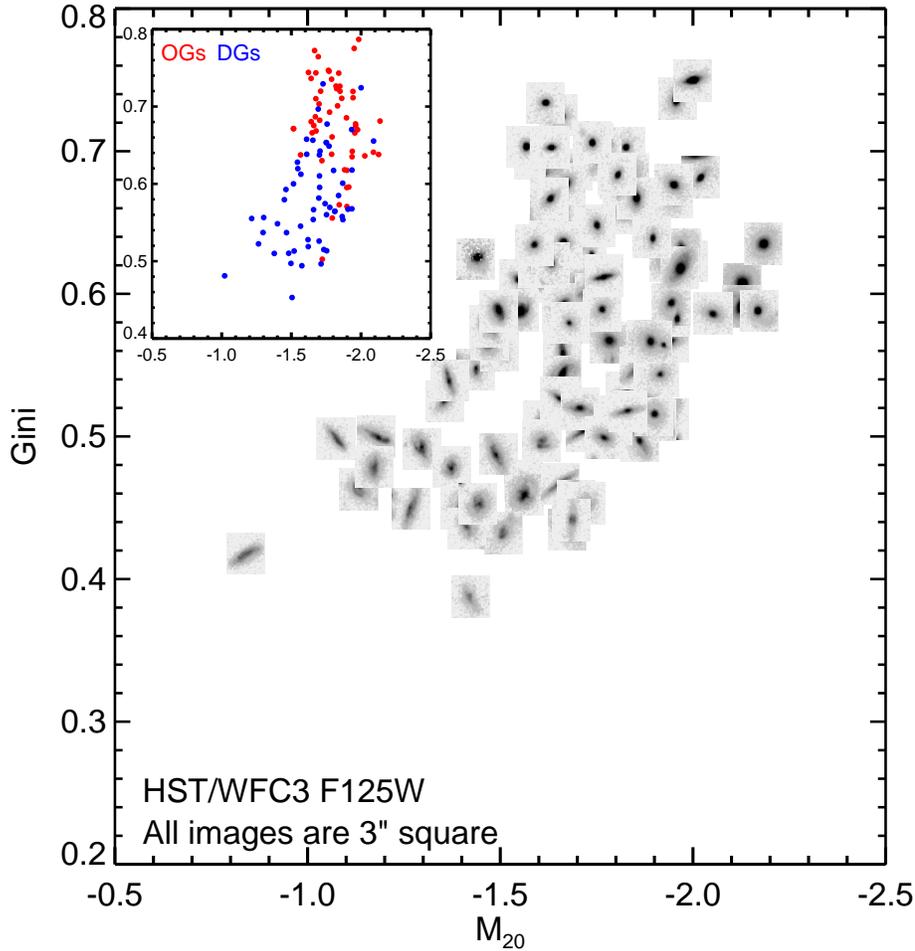}
\caption{Distribution of EROs in the $M_{\rm 20}$ vs. Gini coefficient diagram.
Red and blue solid circles correspond to OGs and DGs, respectively.
 } \label{fig:gm20}
\end{figure*}

\section{Summary}

In this paper, we select a sample of 5241 EROs with $i-K>2.45$
and $K<21.0$ from the catalog of the COSMOS/UltraVISTA field. Based on the UVJ color
criteria, we classify EROs into two main types: old passive galaxies (OGs) and dusty
star-forming galaxies (DGs). In our EROs sample, the fraction of OGs and DGs correspond to
49.7\% and 50.3\%, respectively.

Using the high-resolution ($0''.06~{\rm pixel}^{-1}$) HST/WFC3 F125W imaging, for the first time we
study the morphological and structural properties of OGs and DGs at $\lambda_{\rm rest}\sim6300$~{\AA}.
At a fixed stellar mass, we find that the mean size of OGs (DGs) is smaller by a factor of
$\sim2$ (1.5) than that of local QGs (SFGs) at rest-frame optical wavelength. The average
effective radii of OGs and DGs correspond to $2.09\pm1.13$ kpc ($<z_{\rm p}>=1.31\pm0.23$)
and $3.27\pm1.14$ kpc ($<z_{\rm p}>=1.08\pm0.24$), respectively. Moreover, we also find
that OGs (high $G$ and low $M_{\rm 20}$) in structure show more regular features than
DGs (low $G$ and high $M_{\rm 20}$) at $\lambda_{\rm rest}\sim6300$~{\AA}.
The derived mean ($M_{\rm 20}$, $G$) values of DGs and OGs correspond to ($-$1.65, 0.58)
and ($-$1.81, 0.68), respectively. From the analysis of physical structures of EROs in our
sample, we conclude that OGs and DGs have different evolutionary modes, the structural
features of OGs are consistent with predictions of the hierarchical merging models
(e.g., dry minor merger), while the mass assembly of DGs depends mainly upon the secular
evolution of galaxy.

\normalem
\begin{acknowledgements}
This work is based on observations taken
by the CANDELS Multi-Cycle Treasury Program with the
NASA/ESA HST, which is operated by the Association of
Universities for Research in Astronomy, Inc., under the NASA contract NAS5-26555.
This work is supported by the National Natural Science Foundation of China (NSFC, Nos.
11303002, 11225315, 1320101002, 11433005, and 11421303), the Specialized Research Fund for the
Doctoral Program of Higher Education (SRFDP, No. 20123402110037), the Strategic Priority
Research Program "The Emergence of Cosmological Structures" of the Chinese Academy of
Sciences (No. XDB09000000), the Chinese National 973 Fundamental Science Programs
(973 program) (2015CB857004), the Yunnan Applied Basic Research Projects (2014FB155)
and the Open Research Program of Key Laboratory for Research in Galaxies and Cosmology, CAS.
\end{acknowledgements}

\label{lastpage}
\end{document}